\PassOptionsToPackage{table}{xcolor}
\PassOptionsToPackage{numbers,sort&compress}{natbib}
\documentclass[runningheads]{llncs}

\usepackage[preprint]{colm} 

\setcitestyle{authoryear}

\usepackage{graphicx}
\usepackage{booktabs}
\ifdefined\pdfcompresslevel\else\newcount\pdfcompresslevel\fi
\ifdefined\pdfobjcompresslevel\else\newcount\pdfobjcompresslevel\fi
\ifdefined\pdfoptionpdfminorversion\else\newcount\pdfoptionpdfminorversion\fi
\ifdefined\pdfminorversion\else\newcount\pdfminorversion\fi
\ifdefined\pdfgentounicode\else\newcount\pdfgentounicode\fi
\ifdefined\pdfglyphtounicode\else\def\pdfglyphtounicode#1#2{}\fi
\usepackage[accsupp]{axessibility}
\usepackage{xltabular} 
\usepackage{microtype}
\usepackage{tabularx}
\usepackage{amssymb}
\usepackage{paralist}
\usepackage{todonotes}
\ifdefined\XeTeXversion \else \usepackage{breakurl} \fi
\usepackage{subcaption}
\usepackage{latexsym}
\usepackage{amsmath}
\usepackage{float}
\usepackage{footnote}
\usepackage{enumitem}
\usepackage{bm}
\usepackage{arydshln}
\usepackage{multicol}
\usepackage{multirow}
\usepackage{colortbl}
\usepackage{bbding}
\usepackage{makecell}
\usepackage{mathtools}
\usepackage{imakeidx}
\usepackage{longtable}
\usepackage{tabularx}
\usepackage{wrapfig}
\usepackage{rotating}
\usepackage[edges]{forest}
\usepackage[normalem]{ulem}
\usepackage{CJKutf8}
\ifdefined\XeTeXversion \else \usepackage{awesomebox} \fi
\usepackage[most]{tcolorbox}

\RequirePackage{xspace}
\makeatletter
\DeclareRobustCommand\onedot{\futurelet\@let@token\@onedot}
\def\@onedot{\ifx\@let@token.\else.\null\fi\xspace}

\makeatother

\usepackage{pifont}
\usepackage{lscape}
\usepackage{algorithm}
\usepackage{algpseudocode}

\usepackage{fancyhdr}
\renewcommand{\headrulewidth}{1pt}
\usepackage{fancyvrb}
\usepackage{fvextra}
\usepackage{amsfonts}
\usepackage{wrapfig}

\usepackage{hyperref}
\usepackage{orcidlink}

\definecolor{mydarkblue}{rgb}{0,0.08,0.45}
\definecolor{wkblue}{rgb}{0.2, 0.3, 0.6}
\definecolor{meta-color}{rgb}{0.5, 0.5, 0.5}
\definecolor{darkblue}{rgb}{0, 0, 0.5}
\definecolor{geovistagray}{gray}{0.95}
\definecolor{myblue}{rgb}{0.9, 0.1, 0.94}
\definecolor{mygreen}{rgb}{0.64, 0.56, 0.88}
\definecolor{myyellow}{rgb}{0.68, 0.6, 0.1}
\definecolor{fancygreen}{rgb}{0.33, 0.68, 0.20}
\definecolor{salmon}{rgb}{0.94, 0.52, 0.49}
\definecolor{tablegreen}{rgb}{0.82, 0.94, 0.75}
\definecolor{tableblue}{rgb}{0.81, 0.90, 0.94}
\definecolor{tablered}{rgb}{0.97, 0.85, 0.85}
\definecolor{tableorange}{rgb}{0.96, 0.85, 0.81}
\definecolor{bestcolor}{RGB}{210, 222, 239}
\definecolor{secondcolor}{RGB}{234, 239, 247}
\definecolor{thirdcolor}{RGB}{193, 214, 229}
\definecolor{line-blue}{RGB}{243, 248, 252}
\definecolor{line-green}{RGB}{200,242,200}
\definecolor{line-red}{RGB}{255,215,215}
\definecolor{line-gray}{RGB}{242, 242, 242}
\definecolor{sensepurple}{HTML}{5D2DD6}

\newenvironment{itemize*}%
 {\leftmargini=10pt\begin{itemize}%
  \setlength{\itemsep}{0pt}%
  \setlength{\parskip}{0pt}%
  }%
 {\end{itemize}}
\newenvironment{enumerate*}%
 {\begin{enumerate}%
  \setlength{\itemsep}{0pt}%
  \setlength{\parskip}{0pt}}%
 {\end{enumerate}}

\setcitestyle{numbers,square,comma}

\usepackage{tikz}
\usetikzlibrary{arrows.meta, positioning, fit, backgrounds}
\definecolor{indigo}{RGB}{75,0,130}

\begin{document}

\title{PolyFlow: Continuous Topology Embedding Flow Matching for Artist-style Mesh Generation}



\author{\parbox{\textwidth}{\centering\normalfont
Chunshi Wang$^{1,2,*}$, Haohan Weng$^{2,*}$, Junliang Ye$^{2,3,*}$, Biwen Lei$^{2}$, Yang Li$^{2}$, Zibo Zhao$^{2}$, \\[0.3em]
Zeqiang Lai$^{4,2}$, Kaiyi Zhang$^{5,2}$, Yunhan Yang$^{6,2}$, \\[0.3em]
Zhuo Chen$^{2}$, Chunchao Guo$^{2,\dagger}$, Yawei Luo$^{1,\dagger}$ \\[0.6em]
$^{1}$ZJU \quad $^{2}$Tencent \quad $^{3}$THU \quad $^{4}$CUHK \quad $^{5}$HKUST \quad $^{6}$HKU
}}


\maketitle

\let\thefootnote\relax\footnotetext{
{\small $^{*}$Equal contribution. \quad $^{\dagger}$Corresponding authors: \texttt{chunchaoguo@tencent.com}, \texttt{yaweiluo@zju.edu.cn}}}

\addtocounter{footnote}{0}



\vspace{-1.5cm}
\begin{center}
    \begin{figure}[h!]
      \centering
      \includegraphics[width=\linewidth]{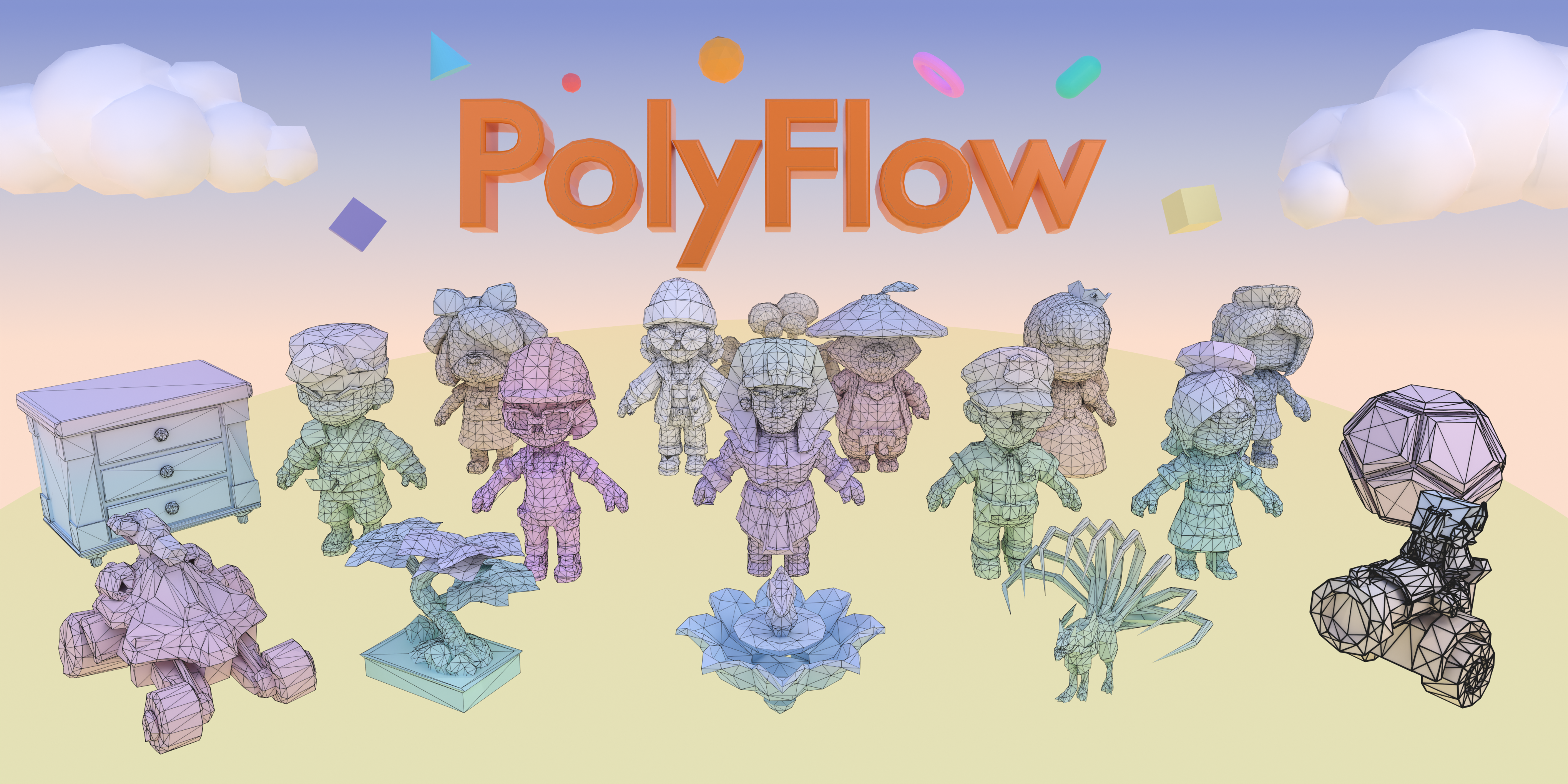}
      \caption{PolyFlow generates meshes with clean, artist-like topology conditioned on point clouds. By denoising vertex positions, normals, and continuous topology embeddings in parallel via flow matching, PolyFlow produces high-quality meshes in seconds with exact vertex-count control.}
      \label{fig:Place}
    \end{figure}
\end{center}


 
\vspace{-1.5cm}

\begin{abstract}
Autoregressive Transformers dominate high-quality mesh generation by producing artist-worthy topologies, yet their inherent sequential decoding induces substantial computational overhead, falling orders of magnitude slower than parallel generative models. On the other hand, while continuous diffusion and flow-matching methods support efficient parallel synthesis across a variety of domains, they cannot be directly applied to meshes: mesh connectivity is inherently discrete and incompatible with standard continuous noise injection and denoising operations. To resolve this fundamental incompatibility, we introduce a compact topology embedder that projects discrete mesh vertex positions and normals into continuous per-vertex embeddings, where the original discrete adjacency information can be faithfully recovered via spacetime distance thresholding. After pretraining and freezing this embedder, any raw mesh can be fully converted into a continuous per-vertex state space unifying position, normal, and implicit topological attributes. Built upon this novel continuous mesh representation, we present PolyFlow, a Transformer-based flow-matching framework that achieves fully parallel vertex state denoising conditioned on extracted point-cloud features. During inference, our model completes generation rapidly via an ODE solver, and supports explicit, precise control over output mesh resolution by directly specifying the target vertex count. Extensive evaluations on the Toys4K benchmark demonstrate that PolyFlow surpasses state-of-the-art autoregressive baselines in both Chamfer Distance and Hausdorff Distance.
\end{abstract}


\section{Introduction}
\label{sec:intro}

Polygonal meshes are the standard surface representation in games, film, and simulation, valued for their explicit connectivity that supports efficient rendering, physical simulation, and artist editing.
While recent advances in feed-forward 3D reconstruction~\cite{instantmesh} can produce dense triangle soups in seconds, converting them into production-ready assets still requires \emph{retopology}, i.e., restructuring the mesh so that its connectivity is clean, purposeful, and controllable in resolution.
Retopology remains one of the most time-consuming steps in 3D content pipelines, motivating a growing body of work on automatic mesh generation with artist-like topology.

The dominant paradigm for automatic retopology is autoregressive (AR) sequence modeling.
These methods serialize a mesh into a one-dimensional token sequence and apply next-token prediction with a Transformer decoder.
PolyGen~\cite{polygen} first demonstrated this idea; MeshGPT~\cite{meshgpt} improved quality with a learned codebook; MeshAnything~\cite{meshanything} and MeshXL~\cite{meshxl} scaled to large datasets; MeshAnythingV2~\cite{meshanythingv2} and EdgeRunner~\cite{edgerunner} introduced more compact tokenizations; BPT~\cite{bpt} achieved a $75\%$ compression ratio, pushing the frontier to meshes exceeding $8{,}000$ faces; and Meshtron~\cite{meshtron} further scaled the model capacity for higher fidelity.

Despite these gradual advances, the sequential nature of AR decoding persists as a fundamental bottleneck: even with the 75\% compression ratio achieved by BPT~\cite{bpt} and the speculative decoding design adopted in XSpecMesh~\cite{xspecmesh}, AR still requires tens of seconds to several minutes to generate a single mesh. This is orders of magnitude slower than diffusion-based image models, which can synthesize high-resolution outputs in under one second.
The excessive latency of AR decoding originates from its rigid sequential dependency: tokens are forced to be generated one at a time. Unlike natural language, which possesses inherent causal ordering, the serialization of mesh vertices and faces is a deliberate design choice of the tokenization pipeline, rather than an intrinsic property of geometric data. Although prior AR mesh generation works have adopted spatial-locality-aware token ordering to mitigate this issue~\cite{bpt,meshanythingv2}, the core limitation of token-wise sequential generation still remains unaddressed.

Meanwhile, flow matching and diffusion transformers have shown that complex structured signals---including high-resolution images~\cite{sd3,dit} and point clouds~\cite{pointe}---can be synthesized in a fully parallel manner with high fidelity and substantially reduced latency. Nevertheless, directly extending these continuous generative frameworks to mesh creation remains non-trivial, owing to the inherently discrete nature of mesh topology. Mesh edges follow a binary existence condition: an edge is either present or absent. AR methods naturally model this binary choice via cross-entropy classification over discrete tokens, whereas continuous diffusion and flow models lack native ability to represent and regularize such categorical structural constraints. \textbf{While vertex coordinates and normals are inherently continuous and well-suited for flow matching formulations, mesh connectivity---the defining characteristic that differentiates meshes from raw point clouds---cannot be easily converted into a continuous representation.}

Our core insight to break this inherent limitation is to circumvent the discrete nature of mesh topology by constructing a novel continuous proxy representation that faithfully approximates discrete vertex adjacency. To this end, we propose a lightweight \emph{topology embedder} that takes ground-truth vertex positions and normals as input and produces a low-dimensional continuous embedding for each vertex. The topology embedder is trained to enforce that the ground-truth discrete adjacency matrix can be recovered from pairwise spacetime distances within the embedding space~\cite{spacemesh}. Once trained and frozen, this embedder transforms the discrete connectivity of arbitrary meshes into a continuous vertex-wise vector field. We concatenate these topological embeddings with vertex positions and normals to formulate a fully continuous per-vertex state $\mathbf{z} = [\mathbf{p},\, \mathbf{n},\, \mathbf{e}]$.
By lifting discrete topological connectivity into a continuous latent space, the complete mesh state becomes fully amenable to flow matching optimization.

Building on this representation, we present \textbf{PolyFlow}, a Transformer-based flow model that denoises the joint state of all vertices in parallel.
Conditioned on point-cloud features from a frozen encoder, PolyFlow generates positions, normals, and topology coordinates simultaneously via an ODE solver, completing inference in seconds.
At the output stage, the generated topology embeddings are decoded back into edges and faces through spacetime distance thresholding.
Beyond the dramatic speedup over AR methods, this parallel formulation brings two additional benefits: it eliminates the serial error accumulation that causes missing semantic parts in AR-generated meshes, and it gives the user direct control over mesh resolution, since the number of vertex tokens is specified as an input, enabling exact vertex-count selection from coarse to fine.

PolyFlow facilitates high-quality retopology conditioned on point clouds, producing meshes with artist-like connectivity and accurate geometry.
Our contributions can be summarized as follows:
\begin{itemize}
    \item We identify the discreteness of mesh topology as the key barrier to applying continuous generative models for retopology, and resolve it by learning a continuous topology embedding supervised via spacetime distance.
    \item We introduce a joint geometry--topology flow state $[\mathbf{p},\,\mathbf{n},\,\mathbf{e}]$ in which vertex positions, normals, and continuous topology coordinates are denoised together by a single Transformer flow model, replacing autoregressive decoding with parallel generation.
    \item We demonstrate that PolyFlow generates meshes with clean topology in seconds with exact vertex-count control, achieving state-of-the-art performance on mesh generation.
\end{itemize}

\section{Related Work}

\subsection{Autoregressive mesh generation.}
The direct generation of native polygon meshes via autoregressive Transformers has rapidly advanced in recent years. PolyGen~\cite{polygen} first demonstrated that meshes can be treated as token sequences, factoring the joint distribution into a vertex model and a face model. MeshGPT~\cite{meshgpt} introduced learned geometric vocabularies via residual vector quantization, compressing face representations and improving topological coherence. Subsequent work focused on more aggressive tokenization: MeshAnything~\cite{meshanything} and MeshAnythingV2~\cite{meshanythingv2} proposed adjacent mesh tokenization (AMT) that halves sequence length by exploiting shared edges, while EdgeRunner~\cite{edgerunner} adapted the classical EdgeBreaker half-edge traversal for neural sequence modeling. BPT~\cite{bpt} achieved a 75\% compression ratio through block-wise indexing and patch aggregation, enabling meshes exceeding 8k faces within standard context windows. PivotMesh~\cite{pivotmesh} introduced hierarchical coarse-to-fine generation via pivot vertices, Meshtron~\cite{meshtron} pushed the boundary to 64k faces using hourglass Transformers with sliding window attention, and QuadGPT~\cite{liu2025quadgptnativequadrilateralmesh} further extended native autoregressive generation from triangle to quadrilateral meshes. Despite these advances, all AR methods share a fundamental constraint: they serialize both continuous coordinates and discrete topology into a single token stream, forcing sequential decoding for geometry that could otherwise be generated in parallel.


\subsection{3D-Native and Flow-Based 3D Generation}
Recent progress in 3D content generation has gradually shifted from optimizing individual 3D assets with external 2D priors to learning 3D-native generative representations. Early text-to-3D methods, pioneered by DreamFusion~\cite{poole2022dreamfusion}, distill visual priors from pretrained 2D diffusion models into 3D representations, enabling open-vocabulary 3D generation without large-scale paired 3D supervision~\cite{lin2023magic3d,chen2023fantasia3d,shi2023mvdream,qiu2024richdreamer,wu2024consistent3d,wang2023prolificdreamer,tang2023dreamgaussian,yi2024gaussiandreamer,ye2024dreamreward,liu2025dreamreward}. While this optimization-based paradigm greatly expands the accessibility of text-to-3D generation, its per-instance optimization process is often computationally expensive and sensitive to multi-view inconsistency. To address these limitations, subsequent studies move toward feed-forward 3D generation by learning compact 3D latent spaces. Representative works such as 3DShape2VecSet~\cite{3dshape2vecset} and TRELLIS~\cite{xiang2025structured} construct 3D-native autoencoding spaces and train generative models directly over structured 3D latents, leading to faster inference and improved geometric fidelity~\cite{zhao2023michelangelo,li2025triposg,he2025sparseflex,li2025sparc3d,li2024craftsman3d,hunyuan3d2025hunyuan3d,lai2025lattice,chen2025ultra3d,ye2025shapellm,UniVerse3D,yang2026physforge,ye2025nano3d}. Beyond holistic shape synthesis, these 3D-native representations also empower part-level understanding and generation, such as native 3D part segmentation~\cite{ma2025p3samnative3dsegmentation}, structure-coherent shape decomposition~\cite{yan2025xparthighfidelitystructure}, and part-aware multimodal modeling~\cite{wang2025partxmllmpartaware3dmultimodal}. These developments suggest that the choice of generative representation is crucial for balancing efficiency, fidelity, and controllability.

Building upon this trend, flow matching~\cite{flowmatching,rectifiedflow} provides a natural framework for modeling continuous 3D states, as it learns velocity fields along simple transport trajectories and can reduce the number of sampling steps compared with conventional diffusion models. Meanwhile, DiT-based architectures~\cite{dit,sit} have shown strong scalability beyond image generation, and Flux-style velocity prediction~\cite{sd3} further demonstrates the effectiveness of flow-based modeling at large scale. Motivated by these advances, our work formulates mesh generation as continuous state transformation rather than discrete token prediction. Specifically, we apply flow matching with velocity prediction to jointly denoise vertex positions, normals, and continuous topology embeddings, treating the full mesh state as a unified continuous representation for parallel generation. This design inherits the efficiency of 3D-native latent modeling while enabling finer geometric control over mesh structure.

\subsection{Mesh connectivity prediction.}
Predicting topology separately from geometry has been explored through several lenses. SpaceMesh~\cite{spacemesh} represents connectivity via continuous per-vertex embeddings under a spacetime distance metric, guaranteeing edge-manifoldness through halfedge cycle construction. DMesh~\cite{dmesh} formulates connectivity as differentiable probabilities over weighted Delaunay triangulations. PointTriNet~\cite{pointtrinet} uses local PointNet classifiers to propose and verify candidate triangles. These methods demonstrate that topology prediction can be decoupled from geometry generation. However, SpaceMesh is limited to $\sim$2k vertices due to transformer memory costs, and Delaunay-based methods cannot represent arbitrary artist-intended tessellations. PolyFlow inherits this continuous spacetime embedding but, rather than predicting connectivity as a separate post-hoc step, integrates it into the generative state so that topology is produced jointly with geometry and recovered in parallel via spacetime distance thresholding.

\section{Methodology}

\begin{figure*}[t]
  \centering
  \includegraphics[width=\linewidth]{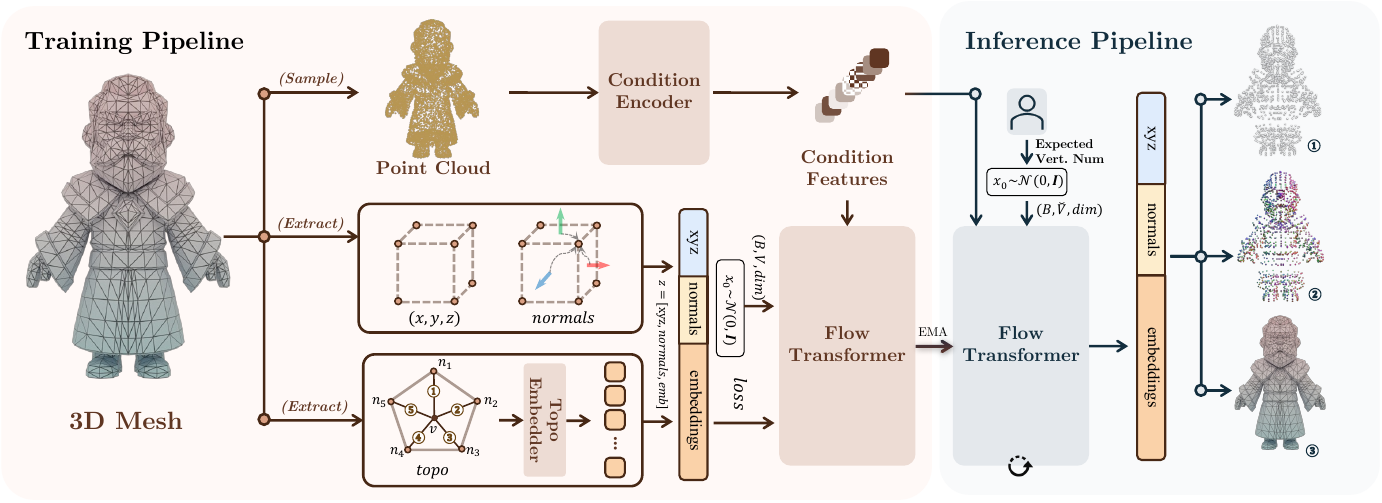}
  \caption{\textbf{Overview of the PolyFlow pipeline.}
  \textbf{Left---Training:} Given a 3D mesh, we sample a point cloud and encode it into condition features via a frozen condition encoder.
  Vertex positions $(x,y,z)$, surface normals, and topology embeddings produced by a frozen topology embedder are concatenated to form the joint flow state $\mathbf{z} = [\mathrm{xyz},\,\mathrm{normals},\,\mathrm{emb}]$ of shape $(B, V, D)$.
  A Flow Transformer is trained to denoise $\mathbf{z}$ from Gaussian noise $\mathbf{x}_0 \sim \mathcal{N}(\mathbf{0}, \mathbf{I})$, conditioned on the point-cloud features.
  \textbf{Right---Inference:} The user specifies an expected vertex count $\hat{V}$; we initialize $\hat{V}$ tokens from noise of shape $(B, \hat{V}, D)$ and denoise them in parallel with the EMA copy of the Flow Transformer.
  The denoised output is split into three channel groups---\ding{192}~vertex positions, \ding{193}~surface normals, and \ding{194}~topology embeddings---from which edges and faces are decoded via spacetime distance thresholding to produce the final mesh.}
  \label{fig:pipeline}
\end{figure*}

We introduce PolyFlow, a flow-matching generative model that denoises native mesh states in parallel.
The method consists of two stages: a \emph{topology embedder} that converts discrete mesh adjacency into continuous per-vertex embeddings (Section~\ref{sec:topo_embedder}), and a Transformer-based flow model that jointly denoises vertex positions, normals, and topology coordinates in a single forward pass (Section~\ref{sec:flow_model}).
An overview of our pipeline is shown in Figure~\ref{fig:pipeline}.

\begin{figure*}[t]
  \centering
  \includegraphics[width=\linewidth]{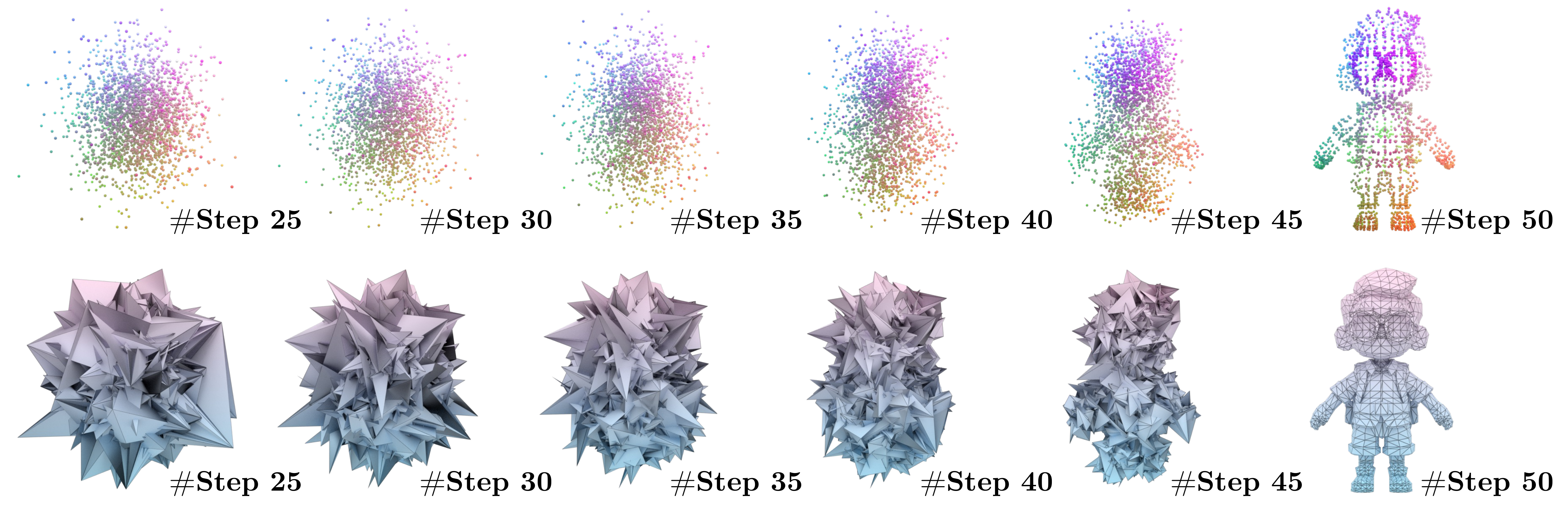}
  \caption{Visualization of the denoising process. Top: vertex positions at selected ODE steps (colored by spatial coordinate). Bottom: meshes decoded from the corresponding topology embeddings. The flow model progressively resolves global shape, local details, and clean connectivity over 50 Euler steps.}
  \label{fig:step_vis}
\end{figure*}

\subsection{Preliminaries}
\label{sec:prelim}

\paragraph{Mesh representation.}
A polygon mesh $\mathcal{M} = (\mathcal{V}, \mathcal{F})$ consists of vertices $\mathcal{V}$ and faces $\mathcal{F}$.
Each face $f_i = (v_{i1}, v_{i2}, v_{i3})$ is defined by three vertices, where each vertex $v_j$ carries a 3D coordinate $(x_j, y_j, z_j)$.
In autoregressive mesh generation~\cite{polygen,meshgpt,bpt}, the mesh is serialized into a one-dimensional token sequence $\mathbf{s} = (s_1, s_2, \dots, s_L)$ and modeled with next-token prediction:
\begin{equation}
  p(\mathbf{s}) = \prod_{i=1}^{L} p(s_i \mid s_{<i}).
\end{equation}
The vertices are typically sorted in z-y-x order and the coordinates are discretized via uniform quantization~\cite{polygen}.
This formulation can produce meshes with clean, artist-like topology, but forces sequential decoding: the model must commit to each token in a fixed serial order, and inference cost scales linearly with sequence length.

\paragraph{Flow matching.}
Flow matching~\cite{flowmatching,rectifiedflow} defines a probability path between a noise distribution $\boldsymbol{\epsilon} \sim \mathcal{N}(\mathbf{0}, \mathbf{I})$ and the data distribution $\mathbf{x} \sim p_{\mathrm{data}}$ via linear interpolation:
\begin{equation}
  \mathbf{z}_t = t\,\mathbf{x} + (1-t)\,\boldsymbol{\epsilon}, \quad t \in [0, 1],
  \label{eq:zt}
\end{equation}
with the conditional velocity field $\mathbf{v} = \mathbf{x} - \boldsymbol{\epsilon}$.
A neural network $\mathbf{v}_\theta$ is trained to regress this velocity by minimizing
\begin{equation}
  \mathcal{L}_{\mathrm{FM}} = \mathbb{E}_{t,\,\mathbf{x},\,\boldsymbol{\epsilon}}\bigl\|\mathbf{v}_\theta(\mathbf{z}_t, t) - \mathbf{v}\bigr\|^2.
  \label{eq:fm_loss}
\end{equation}
At inference time, samples are obtained by solving the ordinary differential equation $\mathrm{d}\mathbf{z}_t / \mathrm{d}t = \mathbf{v}_\theta(\mathbf{z}_t, t)$ from $t{=}0$ to $t{=}1$.

\subsection{Topology Embedder}
\label{sec:topo_embedder}

Mesh connectivity is inherently discrete---an edge either exists or it does not---and cannot be directly noised or denoised by a continuous flow.
To bridge this gap, we train a compact neural network that takes ground-truth vertex positions and normals as input and produces a $d$-dimensional continuous embedding $\mathbf{e}_i \in \mathbb{R}^d$ for each vertex (bottom-left of Figure~\ref{fig:pipeline}).
These embeddings are supervised so that the discrete adjacency matrix can be recovered from pairwise distances in the embedding space.

\paragraph{Spacetime distance.}
Following SpaceMesh~\cite{spacemesh}, we adopt the spacetime distance~\cite{spacemesh} as the pairwise metric.
Each embedding $\mathbf{e}_i$ is split into a space component $\mathbf{e}_i^s \in \mathbb{R}^{d_s}$ and a time component $\mathbf{e}_i^t \in \mathbb{R}^{d_t}$ (with $d_s + d_t = d$):
\begin{equation}
  d^{\mathrm{st}}(\mathbf{e}_i, \mathbf{e}_j) = \|\mathbf{e}_i^s - \mathbf{e}_j^s\|^2 - \|\mathbf{e}_i^t - \mathbf{e}_j^t\|^2.
  \label{eq:spacetime}
\end{equation}
An edge between vertices $i$ and $j$ is predicted when $d^{\mathrm{st}}(\mathbf{e}_i, \mathbf{e}_j) < \tau$, where $\tau$ is a learned threshold.
This pseudo-Riemannian metric has been shown to converge dramatically faster than Euclidean distance for graph edge reconstruction~\cite{spacemesh}.

\paragraph{Edge reconstruction loss.}
The topology embedder is trained with a sampled binary cross-entropy loss over vertex pairs:
\begin{equation}
  \mathcal{L}_{\mathrm{edge}} = \sum_{(i,j) \in \mathcal{E}_{\mathrm{gt}}} \log \sigma\bigl(d^{\mathrm{st}}_{ij} - \tau\bigr) + \lambda \sum_{(i,j) \notin \mathcal{E}_{\mathrm{gt}}} \log \sigma\bigl(\tau - d^{\mathrm{st}}_{ij}\bigr),
  \label{eq:edge_loss}
\end{equation}
where $\sigma$ is the sigmoid function and $\lambda$ balances positive and negative pairs.
Negative pairs are drawn from a mixture of random, spatially-near, and topologically-near (multi-hop) vertex pairs.
Once trained, the topology embedder is frozen and used as a fixed feature extractor throughout Stage~2.

\subsection{Flow Model}
\label{sec:flow_model}

\paragraph{Joint flow state.}
We represent each vertex as a continuous token $\mathbf{z}_i = [\mathbf{p}_i,\, \mathbf{n}_i,\, \mathbf{e}_i] \in \mathbb{R}^{3+3+d}$, concatenating its 3D position $\mathbf{p}_i$, unit surface normal $\mathbf{n}_i$, and the frozen topology embedding $\mathbf{e}_i$ from the topology embedder.
As illustrated in the center of Figure~\ref{fig:pipeline}, a mesh of $V$ vertices is thus a matrix $\mathbf{Z} \in \mathbb{R}^{V \times (3+3+d)}$.
Crucially, the user directly controls $V$: specifying the number of vertex tokens before generation determines the output mesh resolution.

\paragraph{Velocity prediction.}
We apply flow matching (Section~\ref{sec:prelim}) to the joint state $\mathbf{Z}$.
The flow model is a Transformer-based denoiser $\mathbf{v}_\theta$ that takes the noisy state $\mathbf{Z}_t$, the timestep $t$, and a conditioning context $\mathbf{c}$ as input, and predicts the velocity field.
The conditioning context is obtained from a frozen point-cloud encoder that processes the input point cloud.
The training loss is a channel-weighted velocity matching objective:
\begin{equation}
  \mathcal{L} = \sum_{k} w_k \,\mathbb{E}_{t,\,\mathbf{Z},\,\boldsymbol{\epsilon}} \bigl\|\mathbf{v}_\theta^{(k)}(\mathbf{Z}_t, t, \mathbf{c}) - \mathbf{v}^{(k)}\bigr\|^2,
  \label{eq:channel_loss}
\end{equation}
where $k \in \{\mathrm{xyz},\, \mathrm{normal},\, \mathrm{topo}\}$ indexes the channel groups and $w_k$ are per-channel weights that balance the contribution of geometry, normals, and topology.

\subsection{Training}
\label{sec:training}

PolyFlow adopts a two-stage training paradigm.

\paragraph{Stage 1: Topology embedder.}
The topology embedder is trained on the full mesh dataset with the edge reconstruction loss $\mathcal{L}_{\mathrm{edge}}$ (Eq.~\ref{eq:edge_loss}).
After convergence, all parameters are frozen.

\paragraph{Stage 2: Flow model.}
Given a training mesh, the frozen topology embedder produces the per-vertex embedding $\mathbf{e}_i$.
These are concatenated with ground-truth positions and normals to form the clean state $\mathbf{Z} = [\mathbf{P},\, \mathbf{N},\, \mathbf{E}]$, which serves as the target $\mathbf{x}$ in flow matching.
The Transformer denoiser is then trained with the channel-weighted velocity loss (Eq.~\ref{eq:channel_loss}), conditioned on point-cloud features from a frozen encoder with classifier-free guidance dropout.

\subsection{Inference}
\label{sec:inference}

At inference time (right side of Figure~\ref{fig:pipeline}), the user provides a point cloud specifying the desired geometry and selects a target vertex count $\hat{V}$.
We initialize $\hat{V}$ vertex tokens from Gaussian noise $\mathbf{Z}_0 \sim \mathcal{N}(\mathbf{0}, \mathbf{I})$ and solve the ODE $\mathrm{d}\mathbf{Z}_t / \mathrm{d}t = \mathbf{v}_\theta(\mathbf{Z}_t, t, \mathbf{c})$ from $t{=}0$ to $t{=}1$ using an Euler solver with the EMA copy of the trained Flow Transformer.
The denoised state $\mathbf{Z}_1$ is split into three channel groups: \ding{192}~vertex positions $\hat{\mathbf{P}}$, \ding{193}~surface normals $\hat{\mathbf{N}}$, and \ding{194}~topology embeddings $\hat{\mathbf{E}}$.
Because all $\hat{V}$ vertices are denoised in parallel, the ODE solve completes in seconds.

\paragraph{Edge decoding.}
Given the generated topology embeddings $\hat{\mathbf{E}}$, we recover the edge set by evaluating pairwise spacetime distances (Eq.~\ref{eq:spacetime}) over all $\binom{\hat{V}}{2}$ vertex pairs.
An edge is predicted between vertices $i$ and $j$ whenever $d^{\mathrm{st}}(\hat{\mathbf{e}}_i, \hat{\mathbf{e}}_j) < \tau$.
No additional neural network is needed at this stage; the generated embeddings are directly interpretable as spacetime coordinates, and adjacency is recovered purely through distance thresholding.
To handle the quadratic number of pairs efficiently, we evaluate them in batches and retain only those that pass the threshold.

\paragraph{Face extraction.}
Triangular faces are recovered from the decoded edge set by enumerating all 3-cliques: for each edge $(i, j)$, we intersect the neighbor sets of $i$ and $j$, and each common neighbor $k$ with $k > j$ yields a triangle $(i, j, k)$.
This procedure is exact and does not require a separate face prediction network.

\paragraph{Normal-guided winding correction.}
The generated surface normals $\hat{\mathbf{N}}$ serve a second purpose beyond being part of the flow state: they determine the consistent orientation (winding order) of the extracted faces.
For each triangle $(a, b, c)$, we compute the geometric face normal via the cross product $(\hat{\mathbf{p}}_b - \hat{\mathbf{p}}_a) \times (\hat{\mathbf{p}}_c - \hat{\mathbf{p}}_a)$ and compare it against the average of the three generated vertex normals $(\hat{\mathbf{n}}_a + \hat{\mathbf{n}}_b + \hat{\mathbf{n}}_c) / 3$.
If the dot product is negative, the face normal points inward, and we swap two vertex indices to flip the winding.
This ensures a globally consistent outward orientation without requiring a separate post-processing pass such as manifold repair.
Optionally, small boundary holes ($\leq 8$ edges) are filled via ear-clipping triangulation and re-oriented using the same normal-guided procedure.

\section{Experiments}
\label{sec:experiments}

\subsection{Dataset}

PolyFlow is trained on a collection of approximately 5 million meshes assembled from public repositories and licensed 3D assets.
For evaluation, we use the Toys4K dataset~\cite{toys4k}, which contains diverse 3D objects unseen during training.
For each test shape, we uniformly sample a point cloud from its surface as the conditioning input.

\subsection{Baselines and Evaluation Metrics}

We compare against representative mesh generation methods: BPT~\cite{bpt}, MeshAnythingV2~\cite{meshanythingv2}, DeepMesh~\cite{deepmesh}, and FastMesh~\cite{fastmesh}.
For methods with publicly available checkpoints, we use their official released weights for evaluation.
We report Chamfer Distance (CD) and Hausdorff Distance (HD) as geometric fidelity metrics, computed on 1024 points uniformly sampled from the generated and ground-truth mesh surfaces, along with the standard deviation of each metric across test samples.

\subsection{Implementation Details}

\paragraph{Topology embedder.}
The topology embedder takes vertex positions and normals as input and produces a 32-dimensional per-vertex embedding.
The encoder has a hidden dimension of 512 with 12 Transformer layers and 8 attention heads.
The 32-d per-vertex embedding is split equally into space and time components ($d_s{=}d_t{=}16$) under the spacetime distance metric.
The embedder is trained for 350k steps with AdamW ($\beta_1{=}0.9$, $\beta_2{=}0.95$, weight decay $10^{-2}$) at a learning rate of $10^{-4}$ and bf16 mixed precision.
Negative edge samples are drawn from a mixture of random (25\%), spatially-near (25\%), and multi-hop topological neighbors (50\%).
After training, all parameters are frozen.

\paragraph{Flow model.}
The denoiser is a Flux-based DiT~\cite{dit,sd3} with 12 double-stream blocks and 24 single-stream blocks, a hidden size of 768, 16 attention heads, and an MLP ratio of 4.
The input and output dimensionality is 38 (3 xyz + 3 normals + 32 topology embedding).
The conditioning context is provided by a frozen point-cloud VAE encoder from Hunyuan3D-Omni~\cite{hunyuan3d-omni}, which takes 40{,}960 surface points as input and produces 2{,}048 condition tokens of dimension 1024, with a 10\% dropout rate for classifier-free guidance.
Positional encoding is disabled; the model operates in NoPE mode.
The flow model is trained on 64 GPUs with a per-GPU batch size of 1, AdamW ($\beta_1{=}0.9$, $\beta_2{=}0.99$, weight decay $10^{-2}$, $\epsilon{=}10^{-6}$) at a learning rate of $10^{-4}$, bf16 mixed precision, and gradient clipping at norm 1.0.
We maintain an exponential moving average (EMA) of the model weights with a decay of 0.9999.
The channel-weighted velocity loss assigns weights $w_{\mathrm{xyz}}{=}2.0$, $w_{\mathrm{normal}}{=}0.5$, and $w_{\mathrm{topo}}{=}1.0$.
A cosine warmup schedule ramps the learning rate from $10^{-10}$ to its peak over the first 500 steps.

\paragraph{Inference.}
We use the EMA model for inference with an Euler ODE solver and 50 integration steps.
Classifier-free guidance is applied with a scale of 3.0.
Point clouds are sampled at 40{,}960 points from the input surface and fed to the frozen condition encoder.

\subsection{Quantitative Results}

\begin{figure*}[t]
  \centering
  \includegraphics[width=0.90\linewidth]{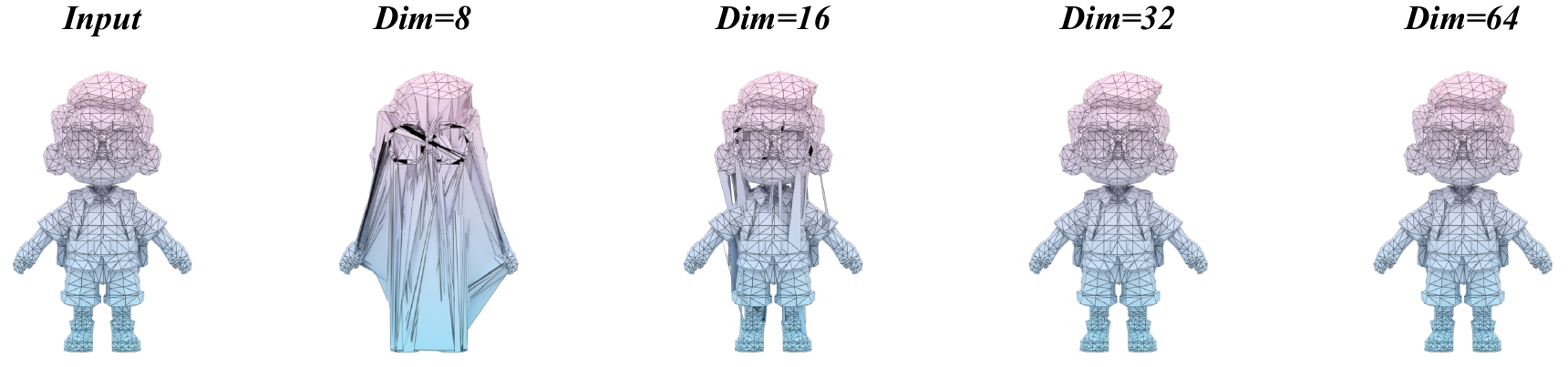}
  \caption{Visual ablation on topology embedding dimension. From left to right: input mesh, and meshes reconstructed by the topology embedder at $d{=}8$, $16$, $32$, and $64$. At $d{=}8$ the embedding space is insufficient to encode adjacency, producing collapsed geometry. At $d{=}32$ the reconstruction is already visually faithful, with no discernible improvement at $d{=}64$.}
  \label{fig:dim_abl}
\end{figure*}

Table~\ref{tab:quantitative} presents the quantitative comparison.
PolyFlow achieves the best performance on both CD and HD, outperforming the strongest AR baseline (BPT) by $43\%$ in CD and $40\%$ in HD.
This improvement is driven by the parallel flow-matching formulation, which jointly denoises geometry and topology without the error accumulation inherent in sequential decoding.
PolyFlow also exhibits the lowest CD standard deviation, indicating stable generation quality across diverse inputs.

\begin{figure*}[t]
  \centering
  \includegraphics[width=\linewidth]{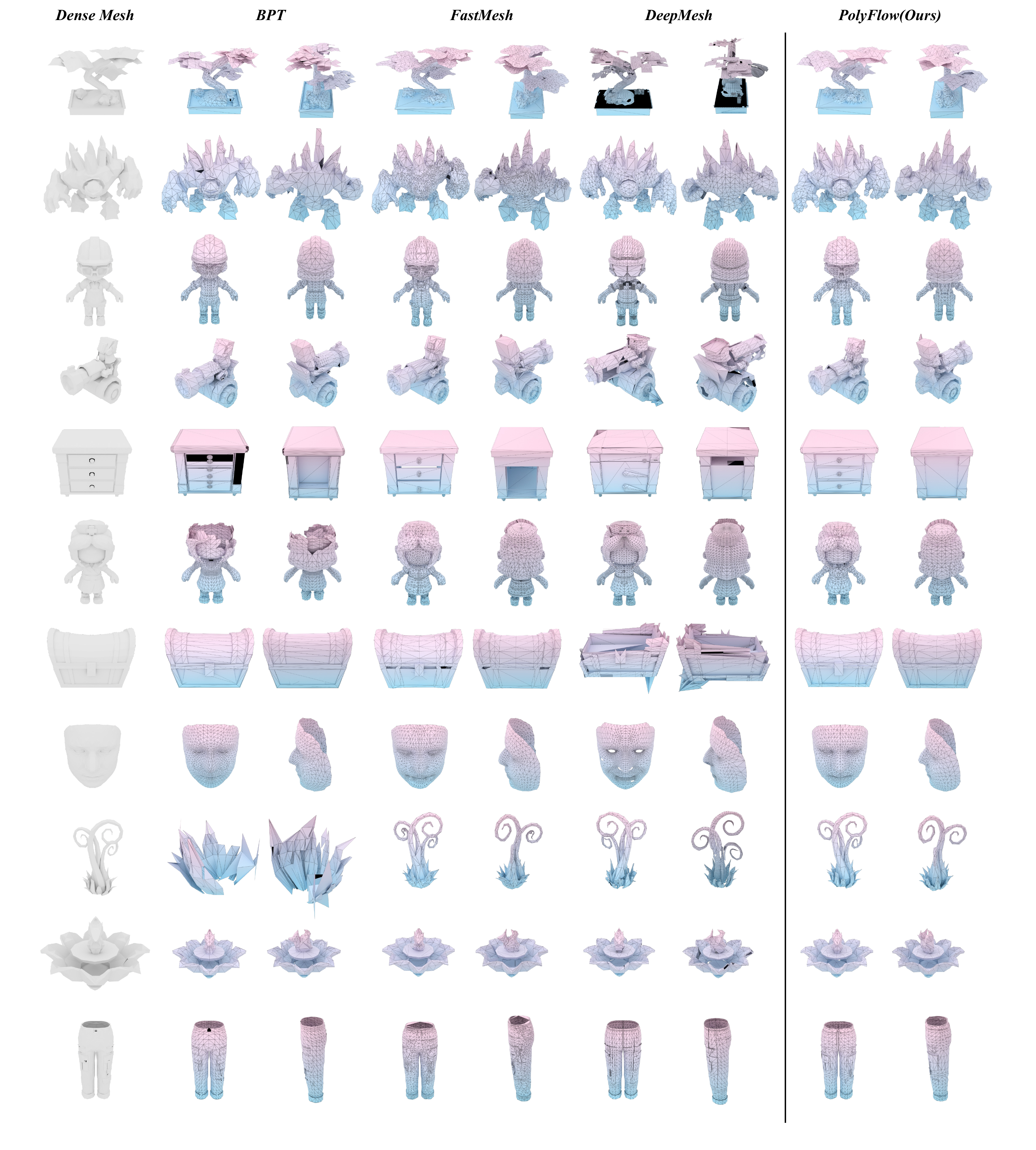}
  \caption{Qualitative comparison of mesh generation results on Toys4K. Each row shows one test shape; from left to right: the input dense mesh, and meshes generated by BPT~\cite{bpt}, FastMesh~\cite{fastmesh}, DeepMesh~\cite{deepmesh}, and PolyFlow (ours). Each method shows two views (front and back). PolyFlow produces meshes with cleaner topology, fewer broken regions, and more faithful geometry across diverse object categories.}
  \label{fig:main_exp}
\end{figure*}

Figure~\ref{fig:main_exp} presents a qualitative comparison across twelve test shapes spanning diverse categories including characters, vehicles, furniture, and organic forms.
BPT and FastMesh frequently produce meshes with broken geometry, missing parts, or distorted proportions, particularly on complex shapes such as the dragon and the spaceship.
DeepMesh, aided by reinforcement learning, improves structural integrity but still exhibits noticeable artifacts in thin structures and fine details.
PolyFlow consistently generates meshes with the most faithful geometry and cleanest topology across all examples, preserving both the global silhouette and local surface details.

\begin{table}[t]
  \caption{Quantitative comparison on point-cloud conditioned mesh generation (Toys4K~\cite{toys4k}). Best results in \textbf{bold}.}
  \label{tab:quantitative}
  \centering
  \small
  \begin{tabular}{lcccc}
    \toprule
    Method & CD $\downarrow$ & HD $\downarrow$ & CD Std & HD Std \\
    \midrule
    MeshAnythingV2 & 0.132 & 0.280 & 0.042 & 0.0548 \\
    FastMesh & 0.130 & 0.271 & 0.037 & 0.0161 \\
    DeepMesh & 0.016 & 0.039 & 0.012 & 0.0196 \\
    BPT & 0.014 & 0.035 & 0.008 & 0.0277 \\
    \midrule
    Ours & \textbf{0.008} & \textbf{0.021} & \textbf{0.001} & \textbf{0.0118} \\
    \bottomrule
  \end{tabular}
\end{table}

\subsection{Qualitative Results}

\begin{figure*}[t]
  \centering
  \begin{subfigure}[t]{\linewidth}
    \centering
    \includegraphics[width=0.90\linewidth]{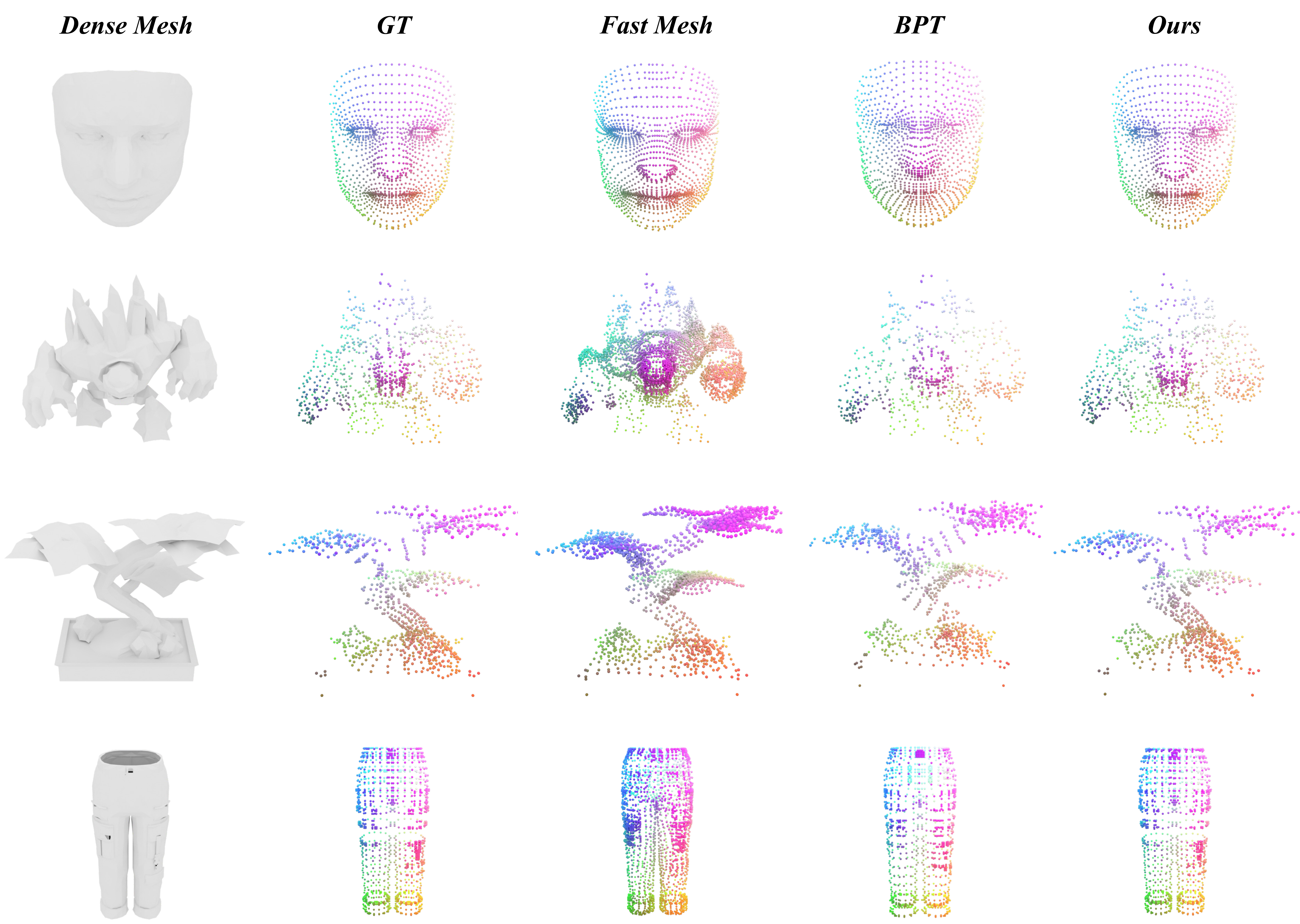}
    \caption{Vertex position comparison. From left to right: the input dense mesh, ground-truth retopology vertices, vertices generated by FastMesh~\cite{fastmesh}, and vertices generated by PolyFlow. Points are colored by spatial position. FastMesh exhibits visible clustering and uneven coverage, whereas PolyFlow produces distributions closely matching the ground truth.}
    \label{fig:points_compare}
  \end{subfigure}
  \par\medskip
  \begin{subfigure}[t]{\linewidth}
    \centering
    \includegraphics[width=0.90\linewidth]{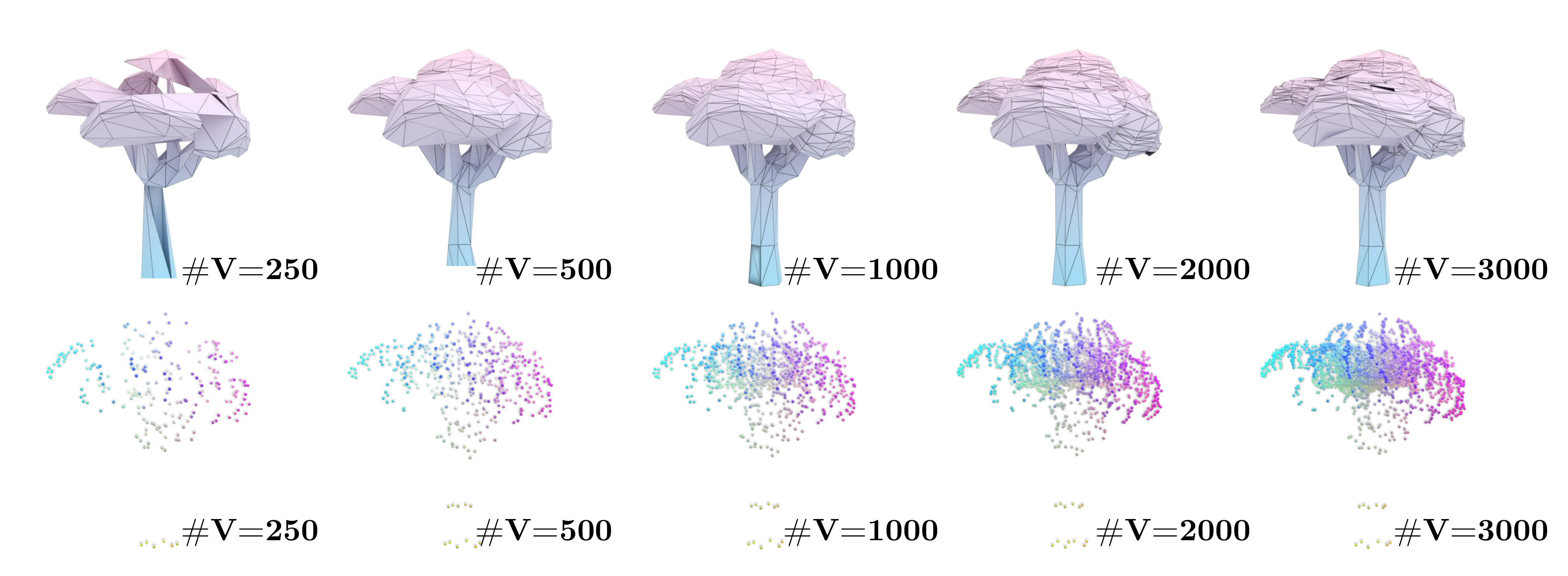}
    \caption{Vertex count control. Given the same point-cloud condition, PolyFlow generates meshes (top row) and the corresponding vertex distributions (bottom row) at user-specified vertex counts from 250 to 3{,}000. The model produces geometrically consistent outputs across the full resolution range, progressively capturing finer details as the vertex budget increases.}
    \label{fig:points_control}
  \end{subfigure}
  \caption{Qualitative evaluation of PolyFlow's vertex generation.}
  \label{fig:qualitative}
\end{figure*}

Figure~\ref{fig:points_compare} compares the generated vertex positions of PolyFlow and FastMesh~\cite{fastmesh} against ground-truth retopology vertices across four test shapes of varying complexity.
FastMesh, which generates vertices through a two-stage autoregressive pipeline, tends to produce uneven point distributions with noticeable clustering artifacts, particularly in geometrically complex regions such as the dragon's extremities and the robot's limbs.
In contrast, PolyFlow's parallel flow-matching formulation produces vertex distributions that closely resemble the ground truth, with uniform spatial coverage and accurate placement along surface features.
This visual difference is consistent with the quantitative gap observed in Table~\ref{tab:quantitative}: the smoother, more faithful vertex placement of PolyFlow directly translates to lower Chamfer and Hausdorff distances.

A distinctive advantage of PolyFlow over autoregressive methods is that the user directly specifies the number of vertices before generation.
Figure~\ref{fig:points_control} demonstrates this capability on a tree model, with the vertex count ranging from 250 to 3{,}000.
At low counts ($\hat{V}{=}250$), the model produces a coarse but recognizable approximation, allocating vertices to the most salient geometric features such as the canopy outline and trunk.
As the count increases, finer details emerge progressively: branches and foliage become increasingly resolved, and the trunk gains smoother curvature.
Throughout the entire range, the overall shape remains geometrically consistent, indicating that the flow model learns a resolution-aware distribution rather than simply scattering additional points.
This controllability is valuable in production pipelines where artists require meshes at multiple levels of detail from a single conditioning input.

Figure~\ref{fig:step_vis} visualizes the denoising trajectory of PolyFlow at selected ODE steps.
The top row shows the vertex positions colored by spatial coordinate, and the bottom row shows the corresponding mesh obtained by decoding the topology embeddings at each step.
At early steps (step 25), the vertices remain dispersed and the decoded mesh is a tangled mass with no recognizable structure.
As the ODE progresses, the point cloud gradually coalesces into a coherent shape: by step 35 the global silhouette emerges, and by step 40 the limbs and head become clearly separated.
In the final steps (45--50), fine details such as facial features and clothing folds are resolved, and the topology embeddings converge to produce a clean, well-connected mesh.
This progressive coarse-to-fine behavior is characteristic of flow-matching generation, where the velocity field first establishes large-scale structure before refining local geometry and connectivity.

\subsection{Ablation Study}
\label{sec:ablation}

\subsubsection{Topology Embedding Dimension}

Table~\ref{tab:topo_dim} ablates the dimensionality of the topology embedding produced by the frozen topology embedder.
We evaluate both edge reconstruction quality (precision, recall, F1) and end-to-end generation performance (CD, HD) of PolyFlow.
At $d{=}8$, the embedding space is too compact to faithfully encode adjacency, resulting in high recall but poor precision.
Increasing the dimension to $d{=}16$ substantially improves precision, while $d{=}32$ achieves near-perfect reconstruction with an F1 of $0.9991$.
Further increasing to $d{=}64$ yields only marginal gains in edge reconstruction ($+0.0005$ in F1) but actually degrades the end-to-end HD (0.025 vs.\ 0.021), likely because the larger flow state increases the learning difficulty for the denoiser without providing meaningful additional topology information.
We therefore adopt $d{=}32$ as the optimal choice that balances topology reconstruction fidelity with downstream generation quality.

Figure~\ref{fig:dim_abl} visualizes the topology embedder's reconstruction at each dimension.
At $d{=}8$, the embedding cannot distinguish nearby vertices, producing a collapsed mesh with severely tangled faces.
At $d{=}16$, the overall shape is recovered but spurious long-range edges remain visible, particularly around the limbs.
At $d{=}32$ and $d{=}64$, the reconstructed connectivity closely matches the input, with $d{=}32$ already producing visually indistinguishable results from $d{=}64$, consistent with the near-identical F1 scores in Table~\ref{tab:topo_dim}.

\begin{table}[t]
  \caption{Ablation on topology embedding dimension. We report edge reconstruction quality of the frozen topology embedder and end-to-end mesh generation performance of PolyFlow.}
  \label{tab:topo_dim}
  \centering
  \begin{tabular}{c|ccc|cc}
    \toprule
    & \multicolumn{3}{c|}{TopoEmbedder} & \multicolumn{2}{c}{PolyFlow} \\
    Dim & Prec $\uparrow$ & Rec $\uparrow$ & F1 $\uparrow$ & CD $\downarrow$ & HD $\downarrow$ \\
    \midrule
    8  & 0.4292 & 0.9969 & 0.6000 & -- & -- \\
    16 & 0.9414 & 0.9996 & 0.9697 & -- & -- \\
    32 & \textbf{0.9983} & \textbf{1.0000} & \textbf{0.9991} & \textbf{0.008} & \textbf{0.021} \\
    64 & 0.9993 & 0.9999 & 0.9996 & 0.007 & 0.025 \\
    \bottomrule
  \end{tabular}
\end{table}

\subsubsection{Inference Speed}

Table~\ref{tab:inference_time} compares the inference time of PolyFlow against BPT~\cite{bpt} and FastMesh~\cite{fastmesh} across varying vertex counts.
BPT, as a purely autoregressive method, exhibits inference time that scales linearly with vertex count, reaching over 9 minutes at 4{,}000 vertices.
FastMesh reduces this substantially through its two-stage pipeline but still requires 36.72s at the same scale.
In contrast, PolyFlow's parallel denoising completes 4{,}000-vertex generation in 5.88s total, achieving a speedup of tens of times over BPT and $6.2\times$ over FastMesh.
The post-processing overhead (spacetime distance decoding and face extraction) remains negligible ($<$70ms) at all scales.

\begin{table}[t]
  \caption{Inference time comparison (seconds). All timings measured on a single NVIDIA A100 GPU.}
  \label{tab:inference_time}
  \centering
  \setlength{\tabcolsep}{3pt}
  \begin{tabular}{c|c|ccc|ccc}
    \toprule
    & BPT$^\dagger$ & \multicolumn{3}{c|}{FastMesh} & \multicolumn{3}{c}{PolyFlow} \\
    \#V & Tot. & S1 & S2 & Tot. & DiT & Post & Tot. \\
    \midrule
     50  & 6.9  & 1.99 & 0.06 & 2.05  & 1.67 & 0.004 & \textbf{1.67} \\
    100  & 13.9 & 2.43 & 0.06 & 2.49  & 1.73 & 0.004 & \textbf{1.73} \\
    500  & 69.3 & 5.90 & 0.11 & 6.01  & 2.04 & 0.006 & \textbf{2.05} \\
   1000  & 138.6 & 10.23 & 0.16 & 10.39 & 2.42 & 0.013 & \textbf{2.43} \\
   1500  & 207.9 & 14.57 & 0.21 & 14.78 & 2.89 & 0.017 & \textbf{2.91} \\
   2500  & 346.6 & 23.24 & 0.32 & 23.56 & 3.89 & 0.030 & \textbf{3.92} \\
   4000  & 554.5 & 36.24 & 0.48 & 36.72 & 5.81 & 0.066 & \textbf{5.88} \\
    \bottomrule
    \multicolumn{8}{l}{\footnotesize $^\dagger$Estimated as (total gen.\ time / total gen.\ vertices) $\times$ target \#V.}
  \end{tabular}
\end{table}

\section{Conclusion}

We have presented PolyFlow, a flow-matching generative model that produces polygonal meshes with clean, artist-like topology by denoising vertex positions, normals, and continuous topology coordinates in parallel.
The key enabler is a topology embedder that converts discrete mesh adjacency into continuous per-vertex embeddings recoverable via spacetime distance, allowing the entire mesh state to be handled by a single Transformer flow model.
Conditioned on point clouds, PolyFlow achieves state-of-the-art geometric fidelity on the Toys4K benchmark while being up to tens of times faster than autoregressive baselines, and provides exact vertex-count control that is unavailable in existing AR methods.


\newpage

\renewcommand{\refname}{References}
\renewcommand{\bibname}{References}
\renewcommand{\bibsection}{\section*{\raggedright \Large References}}
\makeatother

\bibliographystyle{abbrvnat}
\bibliography{egbib}
\end{document}